\documentclass[final]{myias2}

\usepackage{graphicx} 
\usepackage{multirow}
\usepackage{array} 

\usepackage{hyperref} 

\usepackage{subfigure}
\newcommand{\pT}{p$_T$}
\newcommand{\ETmiss}{E$_T^{miss}$}
\RequirePackage{lineno}

\begin{document}

\markboth{Aleandro Nisati, ATLAS Coll.}{Aleandro Nisati, ATLAS Coll.}

\title{Standard Model Higgs boson searches with the ATLAS detector at 
       the Large Hadron Collider}

\author[nisati]{Aleandro Nisati \footnote{on behalf of the ATLAS Collaboration}}
\email{nisati@cern.ch}

\begin{abstract}
The investigation of the mechanism responsible for electroweak
symmetry breaking is one of the most important tasks of the scientific
program of the Large Hadron Collider. The experimental results on the
search of the Standard Model Higgs boson with 1 to 2 fb$^{-1}$ of 
proton proton collision data at $\sqrt s=7$ TeV
recorded by the ATLAS detector are presented and discussed. 
No significant excess of events is found with respect to the expectations from
Standard Model processes, and the production of a Higgs
boson is excluded at 95\% Confidence Level for the mass regions 
144-232, 256-282 and 296-466 GeV.

{\it Proceedings of the Lepton Photon 2011 Conference, to appear in}
Pramana - journal of physics.

\end{abstract}


 
\maketitle


\section{Introduction}

The search for the Standard Model (SM) Higgs boson, is one of the most 
important priorities of the Large Hadron Collider (LHC) scientific
program. A recent review of the theory of the SM Higgs is available here
\cite{djouadi}.
Direct searches at the CERN LEP collider exclude the existence of this 
boson with a mass $m_H$ lower than 114.4 GeV \cite{LEP}.
Searches by the CDF and D0 experiments at the Fermilab $p \bar p$ collider 
have explored the Higgs boson mass up to 200 GeV. The most recent results 
on this search at the Tevatron collider have been presented in this conference
\cite{verzocchi}.

Early results on SM Higgs boson searches recently performed by the
ATLAS and CMS experiments at the LHC, using a data samples of proton-proton
collisions at $\sqrt s=7$ TeV corresponding
to an integrated luminosity from1 to 2 fb$^{-1}$, are also available.

In this paper the results on this search from the ATLAS detector 
\cite{ATLAS} are presented. 
The results on the study performed by CMS are available here
\cite{sharma}.
The channels analysed by ATLAS to search for this boson
are summarized in Table \ref{tab:higgstable}.

\begin{table}[ht]
\begin{center}
\begin{fntable}[0.6\columnwidth]
\begin{tabular}{c|c|c}
\hline
  decay channel              & integrated   & mass range search           \\
                             & luminosity , fb$^{-1}$ & GeV   \\
\hline
  $H\rightarrow\gamma\gamma$ \cite{HggATLAS}  & 1.08 & 110-150 \\
\hline
  $W,ZH\rightarrow l\nu,ll~b \bar b$ \cite{HbbATLAS} & 1.04 & 110-130 \\
\hline
  $H\rightarrow\tau^+\tau^-$  \cite{Htt1ATLAS,Htt2ATLAS} & 1.06 & 110-150 \\
\hline
  $H\rightarrow WW^{(*)}\rightarrow l\nu l\nu$ \cite{HWWlnln} & 1.70 & 110-300\\
\hline
  $H\rightarrow ZZ^{(*)}\rightarrow llll$ \cite{H4lATLAS} & 2.1 & 110-600 \\ 
\hline
  $H\rightarrow ZZ\rightarrow ll\nu\nu$ \cite{HllnnATLAS} & 1.04 & 200-600\\
\hline
  $H\rightarrow ZZ\rightarrow llqq$ \cite{HllqqATLAS} & 1.04 & 200-600 \\
 
\hline
\end{tabular}
\end{fntable}
\caption{The Higgs boson decay channels studied by ATLAS and reported
in this paper (The results on the study of the decay channel 
$H\rightarrow WW^{(*)}\rightarrow l\nu qq$, available in 
\cite{HWWlnqq}, are not included in this paper, nor in the overall
statistical combination described on the last section).
The integrated luminosity related to each of the processes
analysed is also given.
}
\label{tab:higgstable} 
\end{center} 
\end{table}

The following sections describe in some detail the results achieved in 
each of these channels. 
A particular attention is dedicated to the decay channels 
$H\rightarrow\gamma\gamma$, $H\rightarrow WW^{(*)}\rightarrow l\nu\l\nu$
and $H\rightarrow ZZ^{(*)}\rightarrow 4l,ll\nu\nu$ as they play an
important role in setting the overall result of this study.
The analyses take
into account the main systematic experimental effects, such as those
related to the calibration and reconstruction efficiencies of electrons,
muons, tau decays to hadronic final states, jets, transverse missing
energy and b-tagging. Theory uncertainties associated to background
processes as well as to the Higgs boson signal are also taken into
account. The relative uncertainty on the measurement of integrated luminosity,
$\Delta L/L=$ 3.7\%, is also included.
The theory uncertainties on the production cross section and decay branching
fractions of the Higgs boson are taken from \cite{YR1}.

No significant excess of events is found with respect to the expectations 
from Standard Model processes, and hence exclusion limits on the production 
cross section are set at 95\% Confidence Level (C.L.) with a statistical 
analysis based on the {\it CLs} \cite{CLs} method, using the 
{\it profile likelihood ratio} \cite{PL} as test statistic. 
Finally, the statistical combination of the findings from each channel
is presented in discussed in the last section.

\section{$H \rightarrow \gamma \gamma$}

The $H\rightarrow\gamma\gamma$ decay represents one of the few Higgs
processes, whose final state can be fully reconstructed experimentally.
Despite the low branching ratio, BR($H\rightarrow\gamma\gamma)
\approx 2\times 10^{-3}$ for a Higgs boson with mass $m_H=120$, 
this channel provides good experimental sensitivity
in the mass region below 150 GeV \cite{HggATLAS}. The production cross
section of this process is about 0.04 pb.

The background to this final state can be subdivided in two classes:
irreducible and reducible. The irreducible background is made by direct
production of diphoton events from the scattering of the initial proton
constituents, that can be described by Born, bremsstrahlung and box Feymann 
diagrams. The inclusive cross section is about 30 pb. The theoretical
prediction is calculated at the next-to-leading order (NLO) and it is known
with an uncertainty of about 20\%. The reducible background is made by
$\gamma$-jet and jet-jet QCD processes, where one or two jes are misidentified
as photons, respectively.

The excellent performance of the ATLAS electromagnetic calorimeter
is able to ensure a good jet-photon separation, thanks in particular to the
fine granularity along the pseudorapidity $\eta$. This allows to strongly 
reduce the level of contaminations due to the misidentified photon background
arising from jet production.
In the offline reconstruction photons are classified as ``unconverted"
(electromagntic calorimeter clusters with no matching with tracks in the
Inner Detector) and ``converted" (clusters that match with a track pair, or
a single track, consistent with the hypothesis with a photon conversion in
the material of the Inner Detector). 
The single photon reconstruction efficiency is $\approx$ 98\%, while
the identification efficiency ranges typically from 75\% to 90\% for photon
transverse energies between 25 and 100 GeV. 
The purity of the sample (fraction of genuine diphoton events) is about 72\%.
After selection, the fraction of candidates with at least one converted 
photon is 64\%. 

Events with two isolated high-quality photons are selected in the
pseudorapidity range $\eta<1.37$ and $1.52<|\eta|<2.37$, with \pT$>40$ GeV for
the leading photon, and \pT$>25$ GeV the subleading photon.
An accurate mass reconstruction is mandatory to maximize the
signal-to-backgound ratio, and hence have a good sensitivity to $\gamma\gamma$
resonances. The reconstructed width of the Higgs boson peak is expected to
be fully dominated by the experimental resolution. The experimental resolution
is composed by the single-photon energy measurement accuracy, as well as
by the diphoton opening angle in space. For most of the 
$H\rightarrow\gamma\gamma$ topologies, the angular resolution plays an
important role, and therefore it is important to get an accurate measurement
of the photon direction of flight. 
The ATLAS electromagnetic calorimeter allows to 
also measure the photon flight direction with an angular resolution, in
case of unconverted photons, of about
14 mrad at $\eta$=0, a factor $\approx$ 4 smaller than the accuracy 
that is obtained using the average pp interaction point and its spread.
An even better resolution is achieved if at least one of the two
photon is converted. 
This allows to obtain a $\gamma\gamma$ invariant mass resolution
$\delta (m_{\gamma\gamma})\sim$ 1.7 GeV at $m_{\gamma\gamma}$=120 GeV,
dominated by the photon energy resolution.
The analysis is then performed classifying the selected events in five 
complementary categories, depending on whether none, one or both photons
are converted, and on their measured pseudorapidity. For each category, the 
$m_{\gamma\gamma}$ invariant mass distribution is fitted with an
exponential falling distribution to describe the background continuum,
superimposed to a ``crystal-ball" function representing the Higgs boson signal.

\begin{figure}[ht]
\begin{center}
\subfigure[]{
\includegraphics[width=0.434\columnwidth]{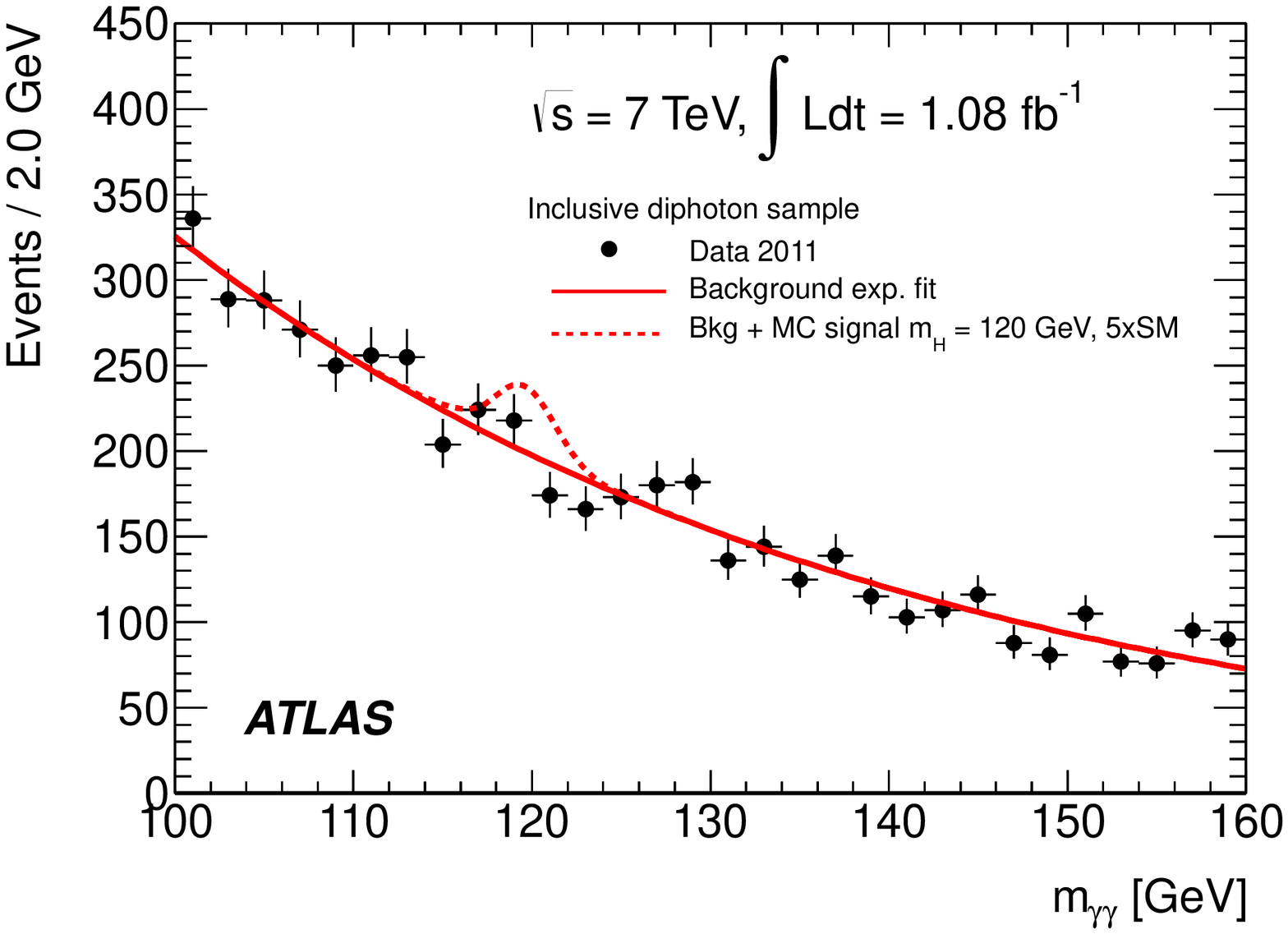}
\label{fig:hgg-mass}
}
\hspace{0.5cm}
\subfigure[]{
\includegraphics[width=0.451\columnwidth]{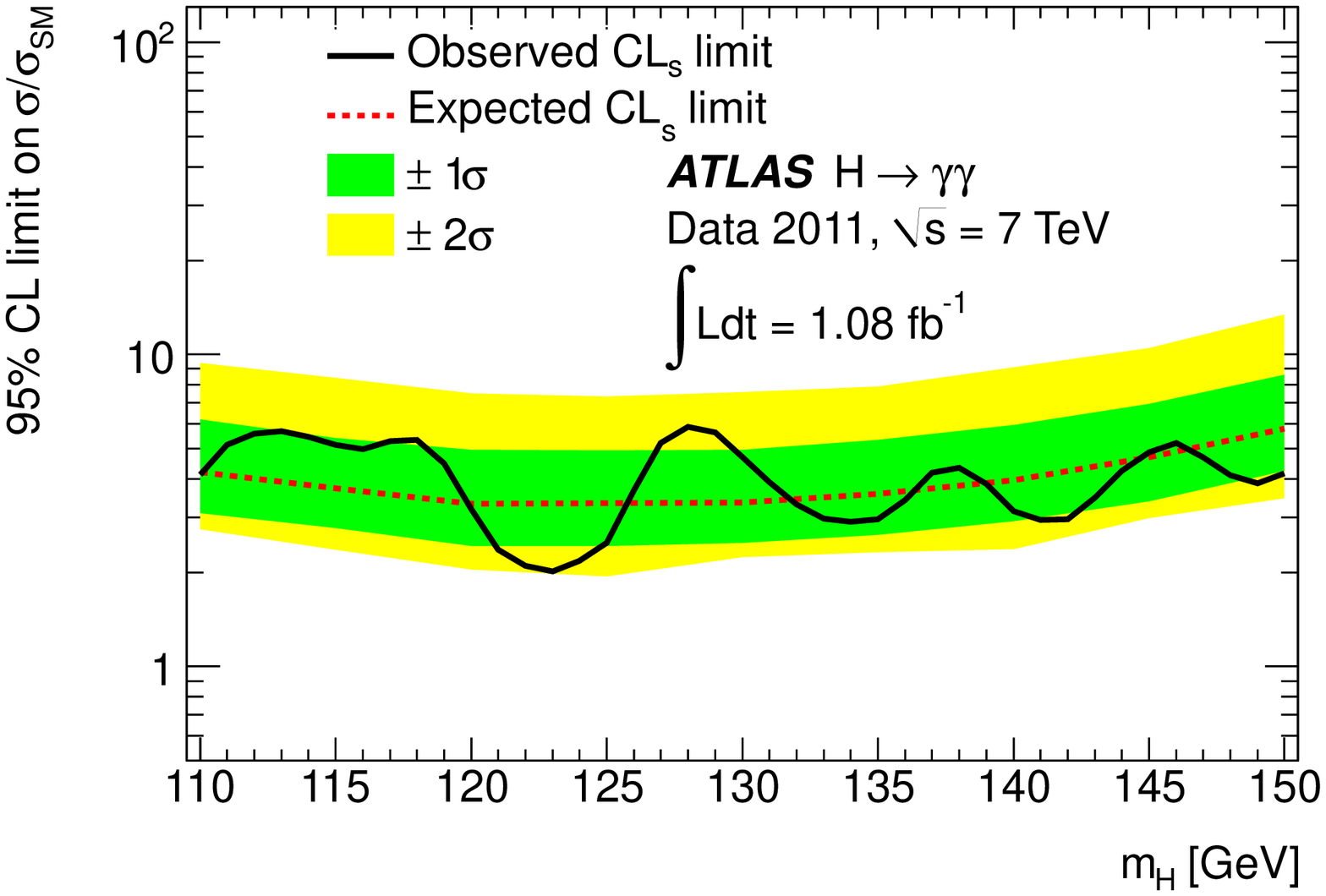}
\label{fig:hgg-limit}
}
\caption{
\subref{fig:hgg-mass} Distribution of the reconstructed invariant mass. 
The exponential fit to the full data sample of the background only hypothesis, 
as well as the expected signal for a Higgs boson of 120 GeV, mass with five 
times the production cross section, are also shown.
\subref{fig:hgg-limit} 95\% confidence level. upper limit 
on a Standard Model-like Higgs boson production cross section, relative 
to the Standard Model cross section, as a fucntion of the Higgs mass hypothesis. 
The solid line describes the
observed limit, while the dashed line shows the median expected limit assuming
no Higgs boson.The bands indicate the expected fluctuation around this limit
as 1 and 2 $\sigma$ level \cite{HggATLAS}.
}
\label{fig:hgg}
\end{center}
\end{figure}

The number of selected events in the 100 - 160 GeV mass range is found to
be 4846, whose measured number of diphoton events is 
3650 $\pm$ 100 $\pm$ 290. The first uncertainty is statistical,
the second is systematic, arising from the estimation made with the 
control region. 
Figure \ref{fig:hgg}\subref{fig:hgg-mass}
 shows the reconstructed 
invariant mass spectrum.  

The total relative systematic uncertainty on the expected signal yield is 12\%, 
dominated by the photon reconstruction and identification efficiency.
The uncertainty on the background measurement is between 5 events at
$m_H=110$ GeV and 3 events at $m_H=150$ GeV, in a mass interval 4 GeV wide.
No evidence of a significant excess of events on top of the background level
is found. Exclusion limits on the inclusive production cross section of a
Standard Model-like Higgs boson relative to the Standard Model 
cross section are derived. 
The result, including systematic uncertainties, are shown in Figure 
\ref{fig:hgg}\subref{fig:hgg-limit}.
The observed limit on the cross section production ranges between
2.0 and 5.8 times those predicted by the SM.
The expected median limit in the case of no signal, varies from 3.3 to
5.8 times the SM predictions.
The expected and the observed limits vary within less then two sigma
in the full mass interval analysed.

\section{W/Z+$H$, $H\rightarrow b\bar b$, and $H\rightarrow\tau^+\tau^-$ }

\begin{figure}[ht]
\begin{center}
\subfigure[]{
\includegraphics[width=0.46\columnwidth]{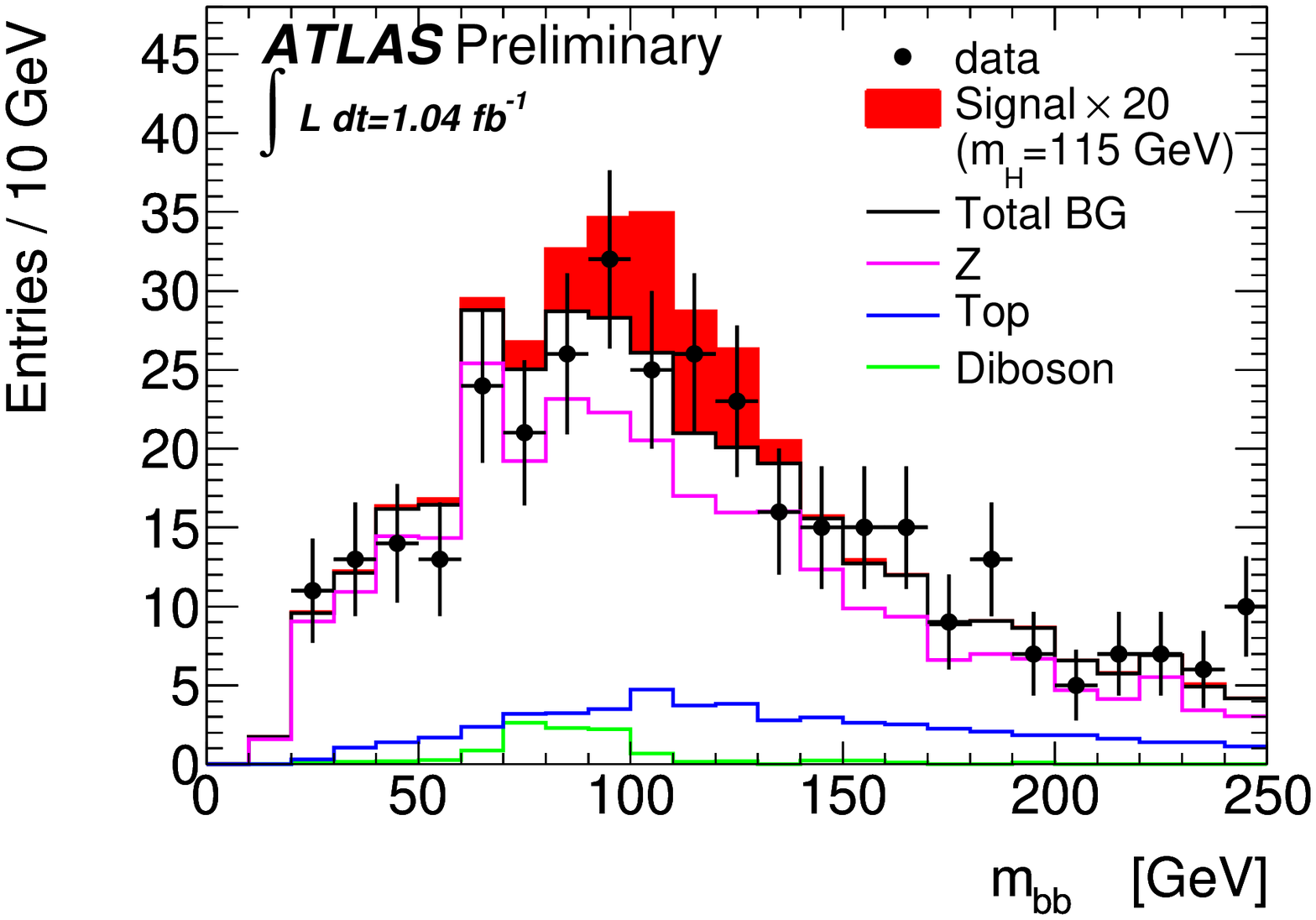}
\label{fig:hbbtt-bb}
}
\hspace{0.5cm}
\subfigure[]{
\includegraphics[width=0.345\columnwidth]{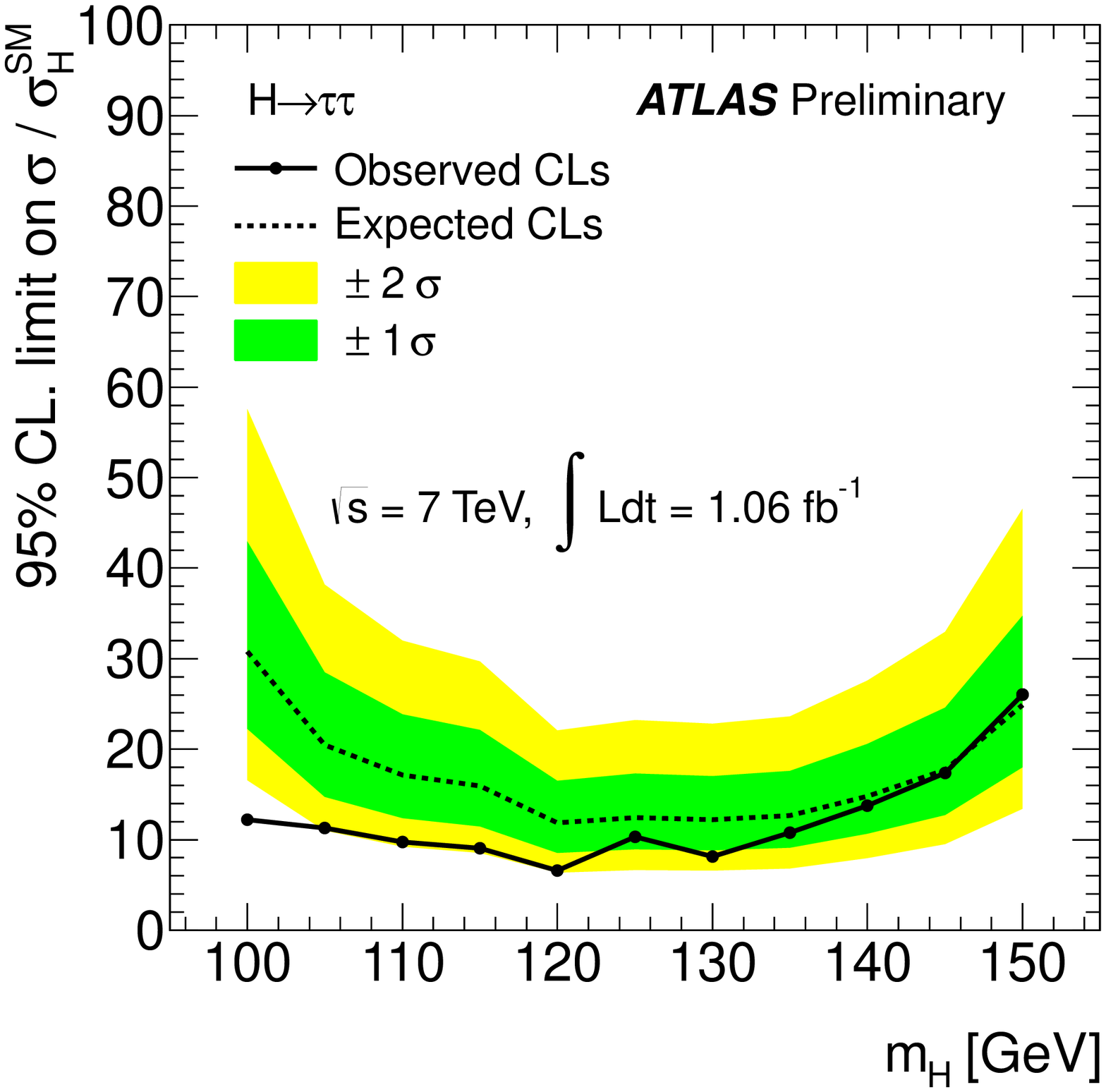}
\label{fig:hbbtt-tt}
}
\caption{
\subref{fig:hbbtt-bb} The invariant mass, $m_{b \bar b}$, for
$ZH\rightarrow llb \bar b$, after full analysis, for $m_H=$115 GeV. 
The signal is enhanced by a factor 20 \cite{HbbATLAS}.  
\subref{fig:hbbtt-tt} Expected and observed limits, at the 95\% confidence
level, on the production of a Standard Model Higgs boson in the 
$\tau^+\tau^-$ final state, as a function of
the mass $m_H$ \cite{Htt1ATLAS,Htt2ATLAS,Hcomboresults}.
}
\label{fig:hbbtt}
\end{center}
\end{figure}

The decay $H\rightarrow b \bar b$ is of particular importance as it is
one of the few channels that offers the possibility of measuring directly the
Higgs couplings to quarks. It also plays an important role in the 
search of this boson in the mass region below 140 GeV. It
is the dominant Higgs decay at low mass, but the QCD jet background makes 
this search in the inclusive channel impossible. On the contrary, it is 
promising in the associated production with gauge bosons W, Z. 
ATLAS has performed an analysis, based on an integrated luminosity of
L=1.04 fb$^{-1}$, searching for $H\rightarrow b \bar b$ produced in
association with a W or a Z. The leptonic decay of the
boson is used both for trigger and offline event preselection. The analysis
proceeds requiring exactly two b-tagged 
jets \cite{btagging} with \pT$> 25$ GeV. 
The $t \bar t$ production and single top are the dominant backgrounds 
in the $WH$ channel;  other important backgrounds are W+jets and
QCD multijet production.
The Z+jets production is the dominant background in the $ZH$ channel; 
other important backgrounds are $t \bar t$ and diboson production.
The study
focuses on the analysis of the jet-jet invariant mass $m_{b \bar b}$, 
where the Higgs signal should appear as a bump on top of the background 
continuum.  
Figure \ref{fig:hbbtt}\subref{fig:hbbtt-bb} shows
the distribution of $m_{b \bar b}$ after all selection cuts, for the
$ZH$ associated production.

In this study the background contributions are either measured directly
from data, or evaluated from Monte Carlo (MC) simulations, and validated with
data control samples.
The total relative uncertainty on the overall background is about 
9\% for both associated production modes.
The systematic uncertainty on the Higgs boson yield is dominated by the
uncertainty on the b-tagging efficiency, about 17\%; the uncertainty
on the {\it Jet Energy Scale} (JES) produces a relative uncertainty 
of about 8\% on the signal yield.

No event excess is found, and 95\% C.L. exclusion
limits are set as a function of the Higgs boson mass. In the mass range
110-130 GeV, a production cross section of about 10-15 times the one
predicted the SM is excluded. 
The obseved limits are within about 1$\sigma$ the expected limits.
More details of this analysis are available here \cite{HbbATLAS}.

The inclusive $H\rightarrow \tau^+\tau^-$ decay is promising for searches 
in the mass range 110$<m_H<$140 GeV. The Vector Boson Fusion process (VBF 
\cite{djouadi}) offers the advantage of a small background, at the price
of a low signal production rate. Three classes of final states are available,
depending on the $\tau$-lepton decay: lepton-lepton ($ll$), lepton-hadron 
($lh$), and hadron-hadron ($hh$).
ATLAS has studied the final states $ll$, produced in association with
at least one high-\pT~ jet, and $lh$ \cite{Htt1ATLAS,Htt2ATLAS}.
The most important backgrounds are represented by the production of
$W,Z\rightarrow$leptons+jets, $Z\rightarrow\tau^+\tau^-$ (largely 
irreducible), dibosons, $t \bar t$, single top and jets. The analysis
is based on the selection of of high-\pT~leptons and of at least one jet.
The {\it collinear approximation} \cite{CollApp} ({\it missing mass calculator}
\cite{mmc}) is adopted to reconstruct the tau momentum 
in the $ll$($lh$) decay mode.
The $\tau\tau$ system invariant mass is studied to search for the Higgs
signal. 

The background contamination in the selected event sample is evaluated 
with methods based on measurements from data, or on predictions from
Monte Carlo simulation.
The overall uncertainty on the background level in the $ll$ final
state is 8\%, while for $lh$  is 19\%.

No event excess is found, and an exclusion limit on the Higgs 
production cross section is set, see Figure 
\ref{fig:hbbtt}\subref{fig:hbbtt-tt}.

\section{$H\rightarrow WW^{(*)}\rightarrow l \nu l \nu$} 

The $H\rightarrow WW^{(*)}\rightarrow l \nu l \nu$ is the most sensitive 
Higgs decay  in the mass range $130 < m_H < 200$ GeV. 
At the same time, this is also one of the most challening channels, 
as the complete reconstruction of the invariant mass of this final state 
is not possible because the presence of neutrinos.
The dominant background originates from the irreducible SM $W^+W^-$
production, but also QCD multijets, W+jets, Drell-Yan and top quark production 
represent important backgrounds to this final state.
The analysis is based on the preselection of two isolated high-\pT~opposite 
sign leptons and large missing transverse energy \ETmiss. 
The leading lepton is required
to have \pT$>$25 GeV, the subleading lepton \pT$>$20 GeV if it is an electron,
15 GeV if it is a muon. QCD and Drell-Yan backgrounds are suppressed by a
\ETmiss~requirement studying the quantity E$_{T,rel}^{miss}$, defined as the 
reconstructed \ETmiss~if the absolute azimuthal angle $\Delta\phi$ between 
the \ETmiss~vector and the nearest lepton (jet) with \pT$>$ 15(25) GeV is 
$\Delta\phi\ge\pi/2$, otherwise 
E$_{T,rel}^{miss}$ = \ETmiss$\cdot\sin\Delta\phi$.
The analysis proceeds classifying the events in two categories: the
0-jet and the 1-jet bins,
depending on whether exactly one jet with \pT$>$25 GeV in the region
$|\eta|<$4.5 has been reconstructed or not. Events with two or more jets
are not considered in this analysis.
In case of the 1-jet bin selection, b-tagged jets are vetoed. The b-tagging
algorithm uses a combination of impact parameter significance and the 
topology of weak \emph{b-} and \emph{c-}hadron decays (the b-tagging
probability of the algorithm used in this analysis to identify \emph{b}-jets
in $t \bar t$ events is 70\%). Additional cuts on the transverse
momentum of the dilepton system, on the jet energy and on the 
missing transverse energy are made, as well as cuts to reject 
Z$\rightarrow\tau\tau$ decays. Topological cuts based on the
measurements of the lepton-lepton invariant mass, transverse momentum and
opening angle in the transverse plane, are also made for both event selections. 
For more details see \cite{HWWlnln}.

\begin{figure}[ht]
\begin{center}
\subfigure[]{
\includegraphics[width=0.4\columnwidth]{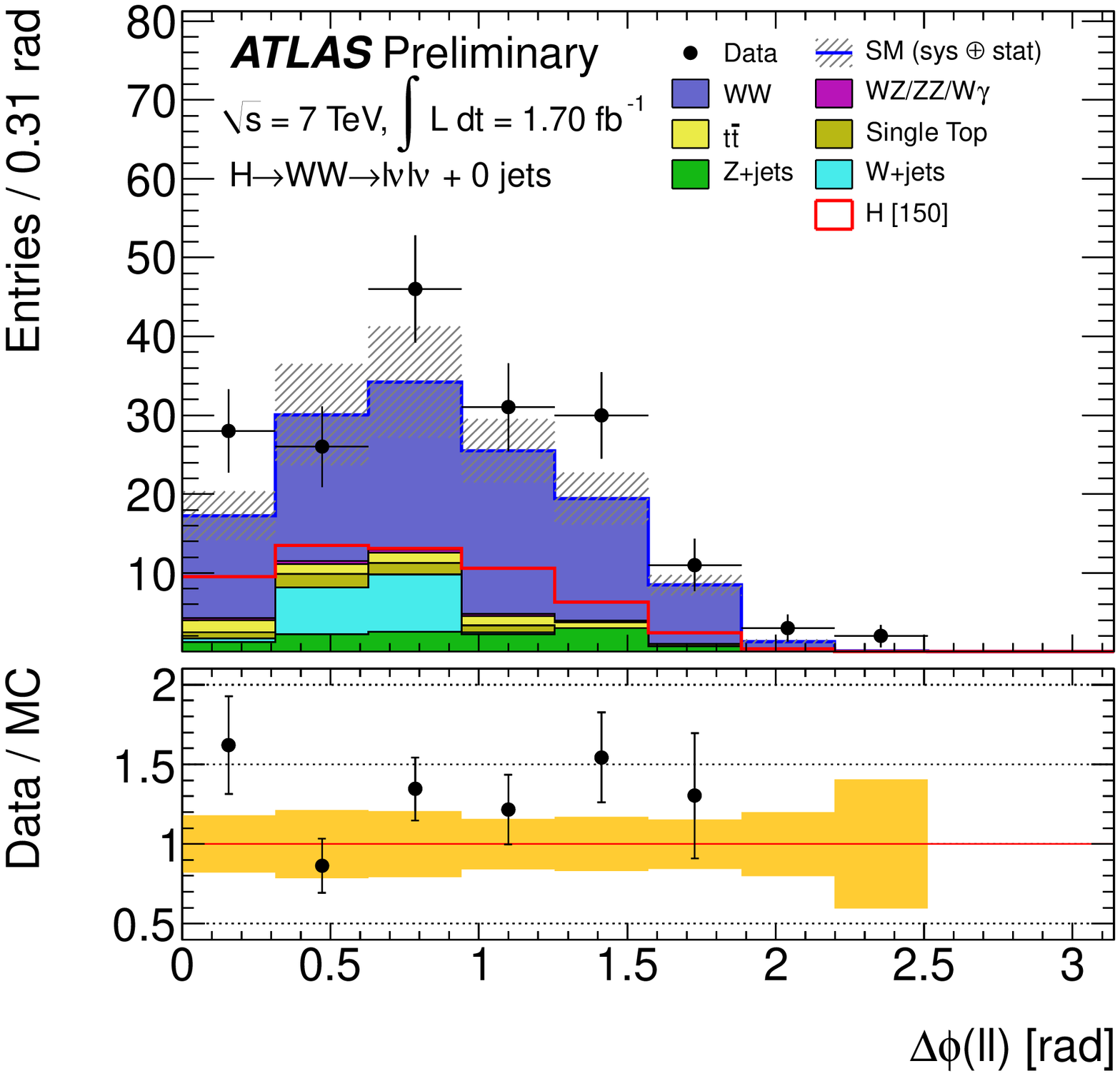}
\label{fig:hwwlnln-dphi0j}
}
\hspace{0.5cm}
\subfigure[]{
\includegraphics[width=0.4\columnwidth]{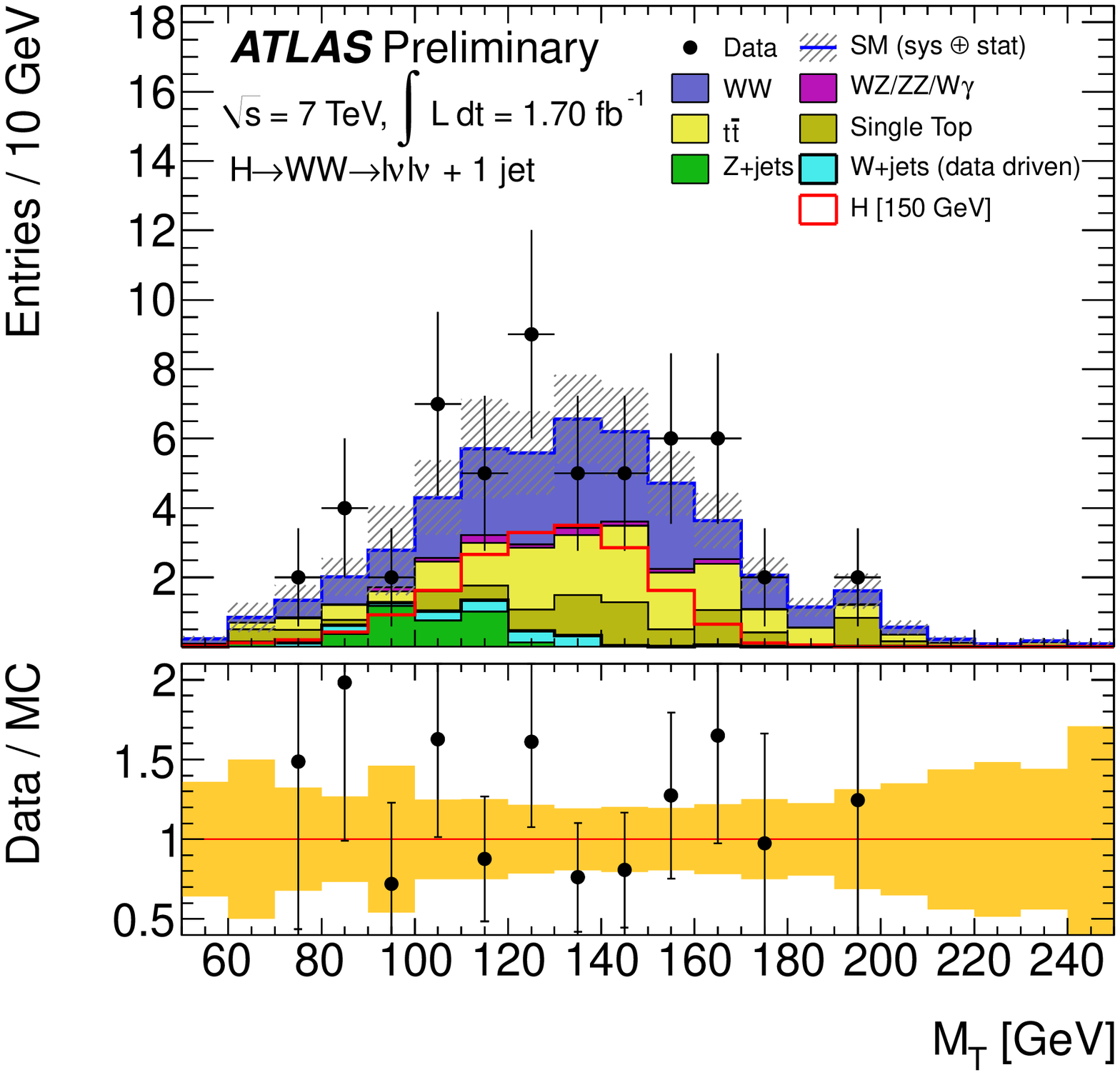}
\label{fig:hwwlnln-mt1j}
}
\caption{
\subref{fig:hwwlnln-dphi0j} $H\rightarrow WW^{(*)}\rightarrow l \nu l \nu$
selection, 0-jet bin: the azimuthal opening angle 
$\Delta\phi_{ll}$ of the two selected leptons after preselection cuts, 
the lepton-lepton transverse momentum and invariant mass cuts. 
\subref{fig:hwwlnln-mt1j} $H\rightarrow WW^{(*)}\rightarrow l \nu l \nu$ 
selection, 1-jet bin: the transverse mass m$_T$ distribution 
after all cuts except for the cut on m$_T$ itself. For both plots 
the expected signal is shown for m$_H$=150 GeV. Also, the lower part
of these plots shows the ratio between the data and the MC prediction. The
(yellow) band indicates the total systematic uncertainty \cite{HWWlnln}. 
}
\label{fig:hwwlnln}
\end{center}
\end{figure}

The ``transverse mass" m$_T$, defined as m$_T$=
$[(E_T^{ll}+E_T^{miss})^2 - (\textbf{P}_T^{ll}+\textbf{P}_T^{miss})^2]^{1/2}$,
 where $\textbf{P}_T^{ll}$ is the transverse momentum of the dilepton system
and $E_T^{ll}$ is its associated energy,
is reconstructed and used for the final selection: for a given Higgs mass
hypothesis m$_H$, events satisfying the condition 
0.75$\cdot$ m$_H < $ m$_T < $m$_H$ are accepted.
Figure \ref{fig:hwwlnln}\subref{fig:hwwlnln-dphi0j} shows for the 0-jet 
selection, the distribution of the lepton-lepton azimuthal opening angle. 
Figure 
\ref{fig:hwwlnln}\subref{fig:hwwlnln-mt1j} shows the $m_T $ distribution 
for the 1-jet selection.

It is important to estimate the background contamination in the 
signal region by measurements performed with data control samples.
In the analysis presented in this document, the three largest background
contributions to this channel, from WW$^{(*)}$, top and W+jets, are evaluated 
using the collision data. For the remaining smaller background contributions,
Monte Carlo predictions are used.
The number of events counted in control regions enriched by a given
SM process are propagated to the signal region through an
extrapolation factor evaluated with MC samples. Extrapolation factors
are evaluated for WW$^{(*)}$ and top processes, while the W+jet contamination
is fully measured in data.
Table \ref{tab:hwwtable} reports the number of event expected from
Standard Model processes, as well as from a Higgs boson with mass
of m$_H$=150 GeV. These numbers are compared with the observed number of
event for each category.
The main uncertainties on the $WW^{(*)}$ and top quark production background
contribution to the selected
events are due to the accuracy of the extrapolation of their measurement
from the control regions to the signal region.
They account for the effects to the the limited knowledge of the JES,
b-tagging efficiency, as well as the renormalization and factorization
scales, and the parton distribution functions associated to the theory models
used to perform these extrapolations. 
The uncertainty on the Higgs boson signal yield is 5\% in the 0-jet
bin and 12\% in the 1-jet bin.

No significant excess of events if found, and therefore an exclusion limit
on the production cross section of a Standard Model-like Higgs boson is
set.

\begin{figure}[ht]
\begin{center}
\includegraphics[width=0.5\columnwidth]{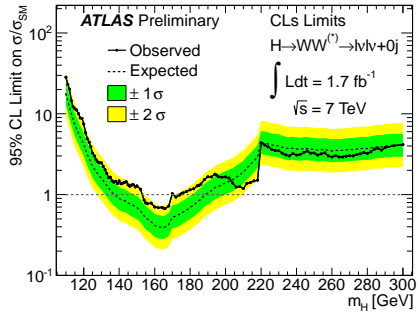}
\caption{The expected (dashed) and observed (solid) 95\% C.L. upper limits on
the production cross section from $H\rightarrow WW^{(*)}\rightarrow l\nu l\nu$
normalized to the Standard Model prediction, as a function of the Higgs boson
mass. The observed limits at neighboring mass points are highly correlated due
to the limited mass resolution of this final state. The bands indicate the
1$\sigma$ and 2$\sigma$ uncertainty regions on the expected limit. The jump
visible at m$_H$=220 GeV is due to the change of the selection criteria at
that mass point \cite{HWWlnln}.}
\label{fig:hww-limit}
\end{center}
\end{figure}

\begin{table}[ht]
\begin{center}
\begin{fntable}[0.6\columnwidth]
\begin{tabular}{c|c|c|c|c|c}
\hline
category & $WW$ & $t \bar t$ & total SM back. & Data & Higgs \\ 
\hline

 0-jet  &  43$\pm$6  &  2.2$\pm$1.4  &  53$\pm$9  &  70 &  34$\pm$7 \\
 1-jet  &  10$\pm$2  &  6.9$\pm$1.9  &  23$\pm$4  &  23 &  12$\pm$3 \\
  
\hline
\end{tabular}
\end{fntable}
\caption{The number of background events, in the 0-jet and 1-jet 
bins, expected from Standard Model processes, observed in the data,
and expected from the production of a Higgs boson with m$_H$ = 150 GeV, using
an integrated luminosity of 1.7 fb$^{-1}$.The uncertainties shown are the
combination of the statistical and all systematic uncertainties, including
the ones from the background extrapolations from control to 
signal regions \cite{HWWlnln}.
}
\label{tab:hwwtable} 
\end{center} 
\end{table}
Figure \ref{fig:hww-limit} shows the observed and expected limits at 95\% C.L. 
for the combined 0-jet and 1-jet analyses as a function of the
Higgs boson mass.
A SM Higgs boson is excluded in the mass interval $154<m_H<186$ GeV,
while the expected  Higgs boson exclusion mass interval is
$135<m_H<196$ GeV. The largest observed difference between observed
and expectd limit is less than 2$\sigma$.

\section{$H\rightarrow ZZ\rightarrow ll \nu\nu$ and $H\rightarrow ZZ\rightarrow llqq$}

The decay $H\rightarrow ZZ\rightarrow ll \nu\nu$ is characterized by a final
state where two leptons are produced in association of large transverse
missing energy. This decay mode offers a significant branching fraction in
combination with a good separation from background processes. These
processes are mainly the irreducible diboson production WW, WZ and ZZ,
plus the reducible background processes  represented by 
top quark final states, as well as W,Z+jets.
The analysis proceeds with the selection of two isolated same-flavour
opposite charge high-\pT~leptons whose invariant mass is consistent with
the Z-mass, and large \ETmiss. Toplogical cuts are applied to suppress the
W,Z+jets background. Finally the transverse mass distribution $m_T$,
defined as:

\begin{equation}
m_T^2 =
\left(\sqrt{m^2_Z+|\bf{p}_T^{ll}|^2}+\sqrt{m^2_Z+|\bf{p}_T^{miss}|^2}\right)^2 + (\bf{p}_T^{ll}+\bf{p}_T^{miss})^2
\end{equation}

is studied to search for contributions from the production and
decay of the SM Higgs boson ($\bf{p}_T^{miss}$ is the missing transverse
energy vector).

The systematic uncertainties include experimental uncertainties related to
the calibration and reconstruction efficiencies of electrons, muons, jets,
and b-jets that are propagated also the reconstruction of \ETmiss.
Uncertainties relative to the Higgs signal yield prediction (about 12\%
for the gluon-gluon production and 4\% for the VBF)
as well as to the ZZ background (11\%), have been also taken into account.
  
\begin{figure}[ht]
\begin{center}
\subfigure[]{
\includegraphics[width=0.44\columnwidth]{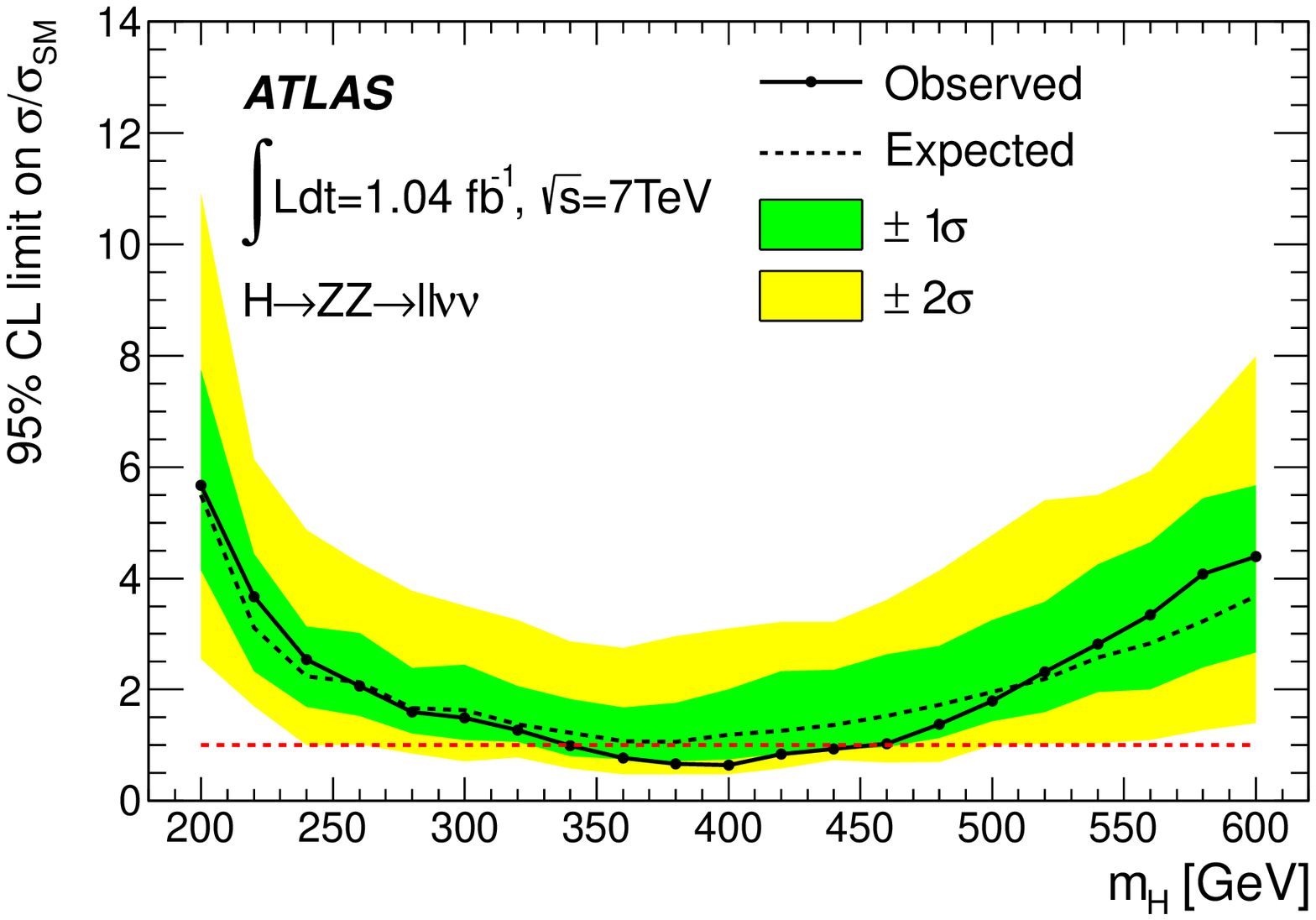}
\label{fig:hzzllnnqq-llnn}
}
\hspace{0.5cm}
\subfigure[]{
\includegraphics[width=0.41\columnwidth]{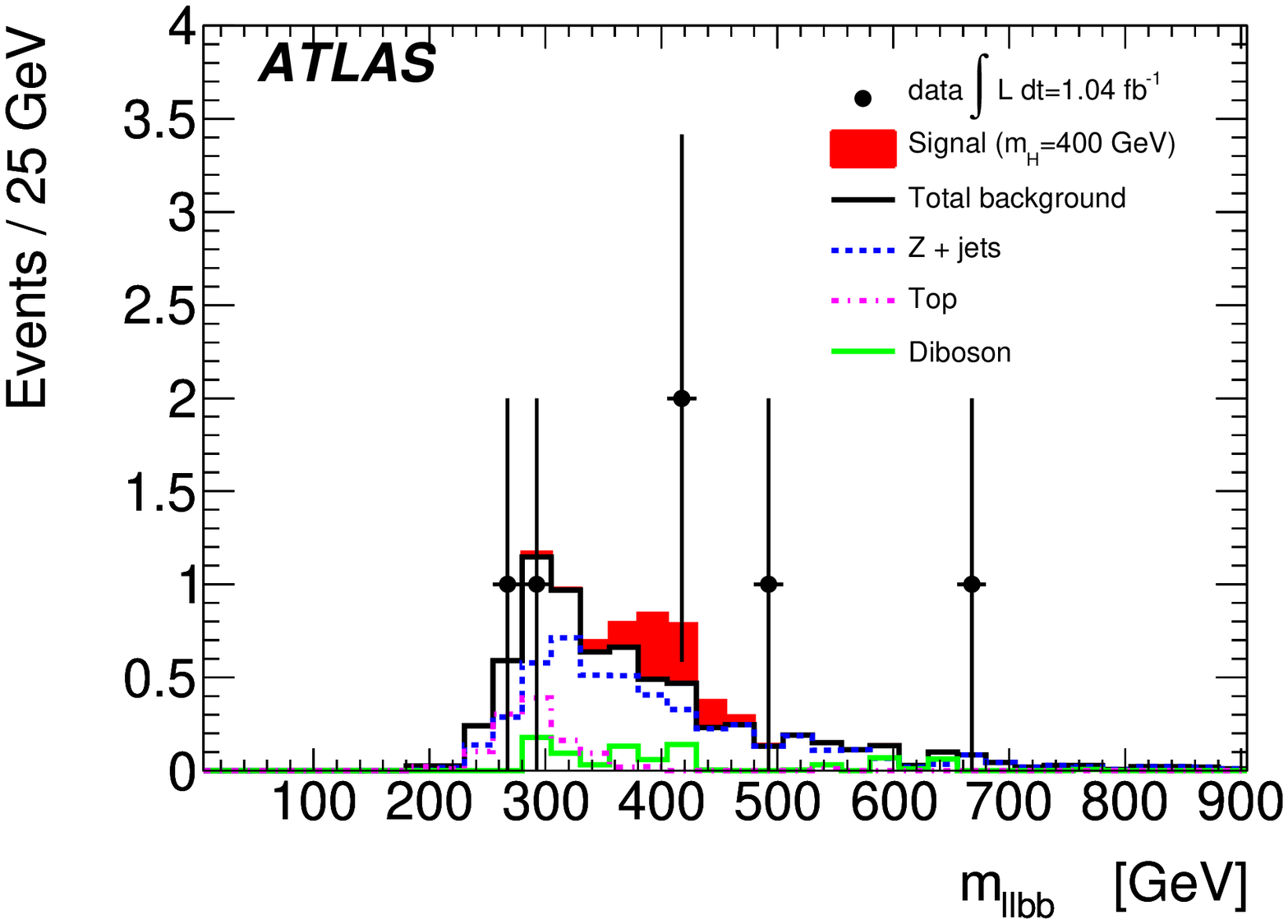}
\label{fig:hzzllnnqq-llqq}
}
\caption{
\subref{fig:hzzllnnqq-llnn} Observed (full line) and expected 
(dashed line) 95\% confidence level upper limits on the Higgs boson 
production cross section divided by the SM prediction for the 
$H\rightarrow ll\nu\nu$ search.
The green and yellow bands indicate the 1 $\sigma$ and 2 $\sigma$ fluctuations,
respectively, around the median sensitivity \cite{HllnnATLAS}.
\subref{fig:hzzllnnqq-llqq} The invariant mass of the $lljj$ system for 
the channel obtained by the $H\rightarrow ZZ\rightarrow llqq$ search analysis.
The expected Higgs~boson signal for $m_H=400$ GeV is also shown
\cite{HllqqATLAS}.
}
\label{fig:hzzllnnqq}
\end{center}
\end{figure}

The observed $m_T$ distribution is consistent with the expectations from 
Standard Model processes. Exclusion limits at 95\% C.L. have been set as 
a function of the Higgs boson mass. The results are shown in Figure 
\ref{fig:hzzllnnqq}\subref{fig:hzzllnnqq-llnn}. 
A Standard Model Higgs boson with mass in the region $340<m_H<450$ GeV
is excluded. The expected limit is lowest around $m_H=380$ GeV where it is 
1.1 times the cross section predicted by the SM. 
The observed and expcted limits agree within 2$\sigma$ over the entire mass range.
The details of this analysis are in \cite{HllnnATLAS}.

Another channel important for high-mass SM Higgs boson searches is the
decay $H\rightarrow ZZ\rightarrow llqq$, also characterized by a large
branching fraction. The event selection is based on the reconstruction of
final states with two same flavour opposite charge high-\pT~ leptons,
whose invariant mass is consistent with the Z boson mass $m_Z$, 
in association with two high-\pT ~jets again with an invariant mass 
consistent with $m_Z$, and no significant transverse missing energy.
The distribution studied for the Higgs boson search is obviously the
two-lepton and two-jet system invariant mass.
This distribution is studied also for events with both jets 
b-tagged \cite{btagging}.
The dominant background is the irreducible ZZ production, the reducible
WZ, Z,W+jets and QCD jets. 
The Higgs boson signal would appear as a bump on top of the expected background,
see Figure \ref{fig:hzzllnnqq}\subref{fig:hzzllnnqq-llqq}.
No significant excess is observed, and therefore 95\% C.L. 
production cross section exclusion limits are set.
A production cross section about 1.7 times the one predicted by the SM
is excluded for $m_H\sim 360$ GeV. The expected limit is between 2.7 and
9 the SM cross section in the mass range 200 to 600 GeV.
The details of this analysis are in \cite{HllqqATLAS}.

\section{H$\rightarrow ZZ^{(*)}\rightarrow llll$}

Three distinct channels are studied in this analysis: 
$H\rightarrow ZZ^{(*)}
\rightarrow e^+e^-e^+e^-$, 
\newline
$\mu^+\mu^-\mu^+\mu^-$, $e^+e^-\mu^+\mu^-$. The 
production of SM $ZZ^{(*)}$ dibosons represents the irreducible background.
Processes such as $Z+jets$ (in particular $Zb \bar b$) and
$t \bar t$, where at least two fake/non-prompt leptons are reconstructed 
from jets or heavy quark semi-leptonic decays, represent the more important 
reducible backgrounds to these final states.
Data have been selected using single-lepton triggers. In the offline analysis,
events with two pairs of same-flavour opposite-sign isolated leptons are
selected. Each lepton must have \pT$>7$ GeV;  
electrons (muons) should lie in the region $|\eta|<2.47(2.50)$. The electron
\pT~threshold is increased to 15 GeV if the associated pseudorapidity is
between $1.37<|\eta|<1.52)$. At least two leptons must have \pT$>20$ GeV.
The two lowest \pT~leptons in events with the 4-lepton invariant mass
of $m_{4l}<190$ GeV are required to be tightly associated to the primary
vertex with cuts to the impact parameter significance (defined  
as the transverse impact parameter to the primary vertex divided by its
uncertainty).

The invariant mass $m_{12}$ of the lepton pair closest to the Z mass 
must fulfill $|m_{12}-m_Z|<15$ GeV. The invariant mass $m_{34}$ of the
other pair is required to be lower than 115 GeV and greater than a threshold
depending on the reconstructed 4-lepton mass. The full-width half-maximum
(FWHM) of the reconstructed Higgs particle, expected to be dominated by 
the experimental mass resolution for low $m_H$ values, 
varies between 4.5 (4$\mu$) GeV and 6.5 GeV (4$e$) for $m_H=130$ GeV. 
At high $m_H$ the reconstructed width is dominated by the intrinsic
width (e.g. FWHM=35 GeV for $m_H$=400 GeV).
The expected contribution from SM background is estimated with Monte Carlo
simulation, except the Z+jets process whose contribution is measured in data.
The irreducible ZZ$^{(*)}$ process is predicted with an uncertainty of
15\% (includes both quark-quark, quark-gluon and gluon-gluon scattering),
while the top estimation, predicted with an uncertainty of 10\%, is validated
with data control samples. 
The Z+jet(s) contribution is evaluated starting from a sample of Z+lepton-pair
selected with no isolation or impact parameter requirements. This yield is then
extrapolated to the signal region applying reduction factors estimated
with MC simulation. The uncertainty on the event contamination from this source
varies in the range 20\% to 40\%.

\begin{figure}[ht]
\begin{center}
\subfigure[]{
\includegraphics[width=0.4\columnwidth]{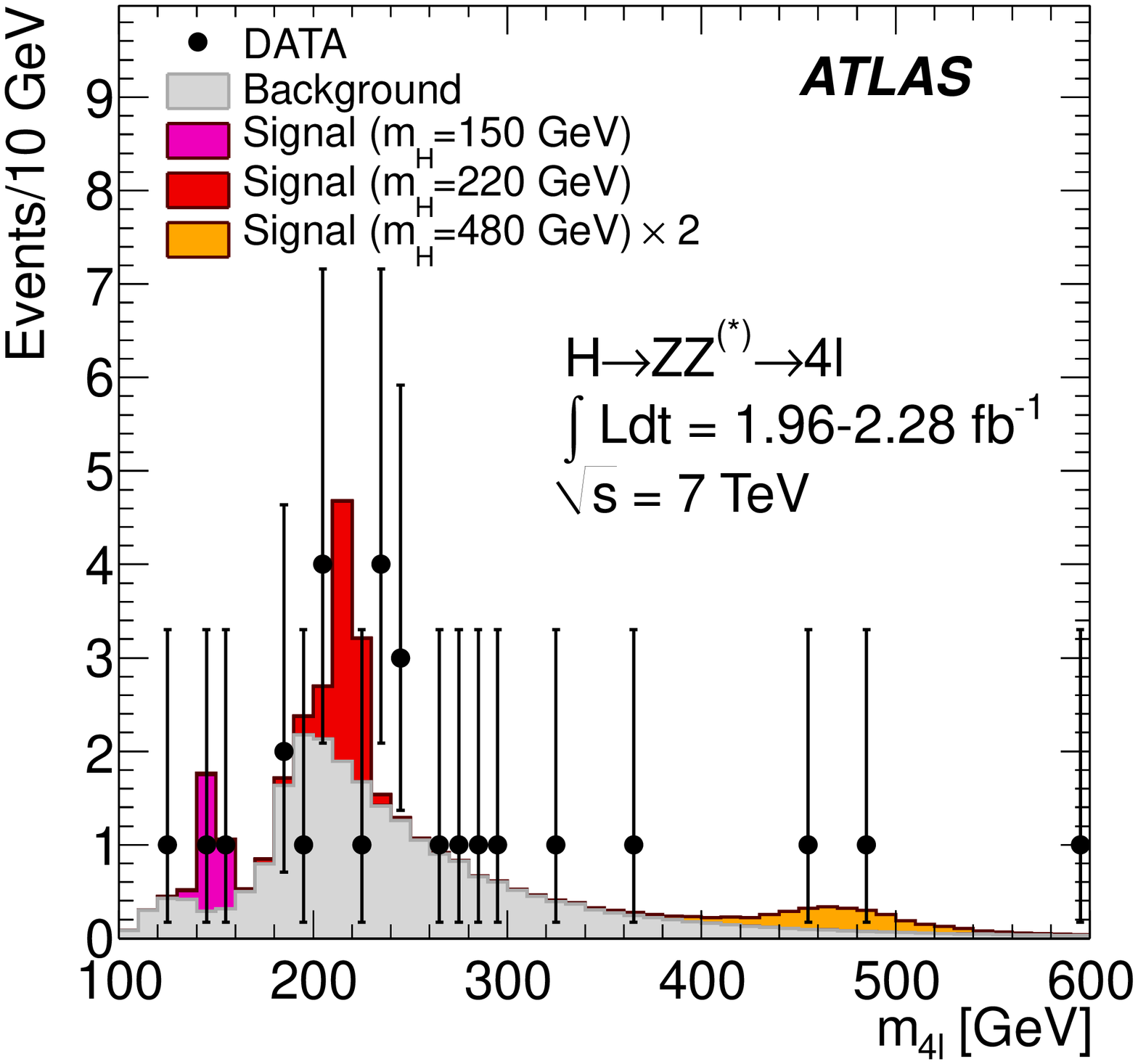}
\label{fig:hzz4l-invm}
}
\hspace{0.5cm}
\subfigure[]{
\includegraphics[width=0.4\columnwidth]{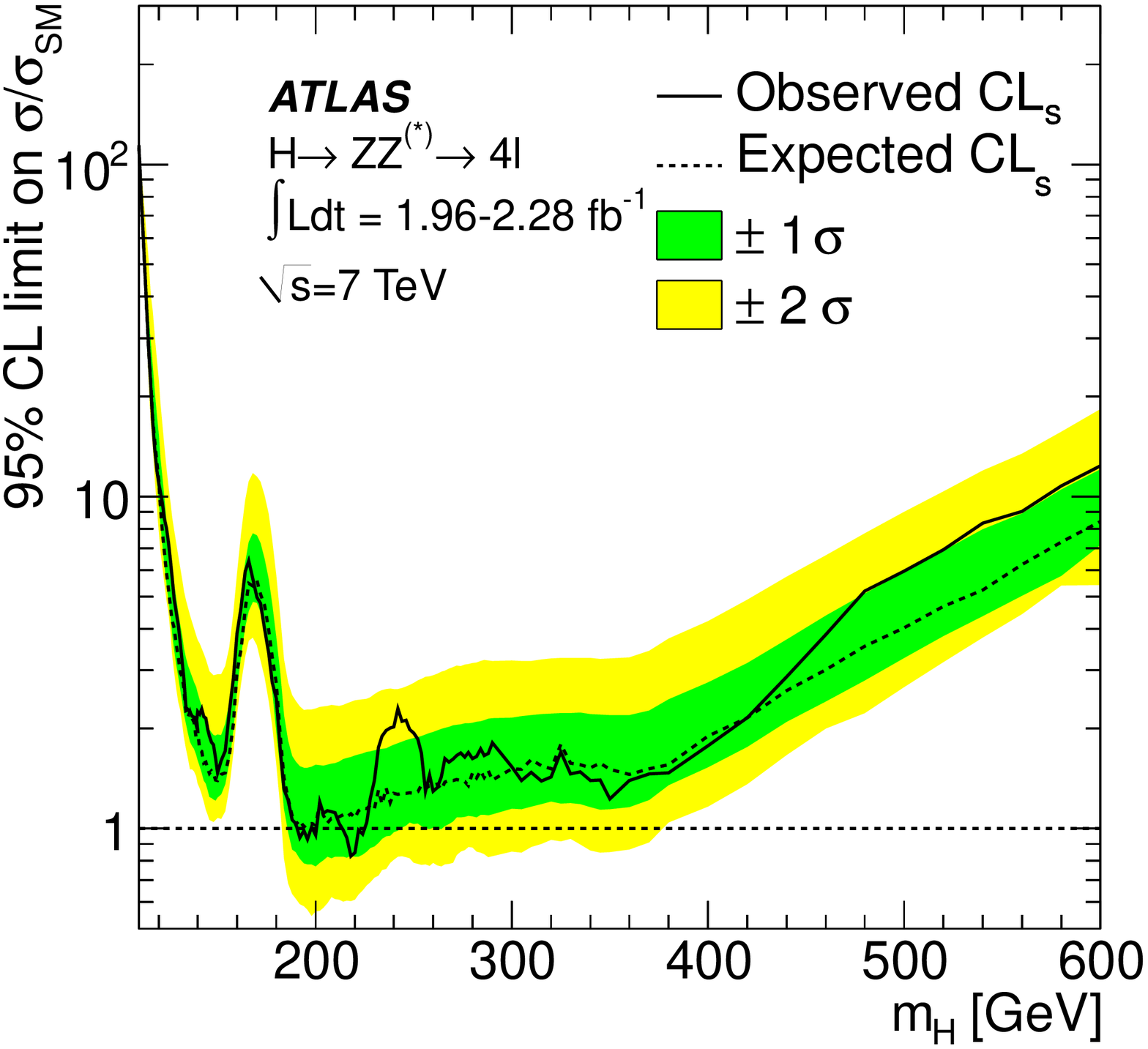}
\label{fig:hzz4l-limit}
}
\caption{
\subref{fig:hzz4l-invm} Invariant mass distribution for the selected
4-lepton event candidates.The data are compared to the background
expectations. 
\subref{fig:hzz4l-limit} The expected (dashed) and observed (full line)
95\% C.L. upper limits on the SM Higgs boson production cross section,
normalized to the SM prediction, as a function of its mass. Green and yellow
bands indicate the fluctuations at 1$\sigma$ and 2$\sigma$ around the 
value of the median \cite{H4lATLAS}.
}
\label{fig:hzz4l}
\end{center}
\end{figure} 

Figure \ref{fig:hzz4l}\subref{fig:hzz4l-invm} shows the distribution of the 
invariant mass of the selected candidates, compared to the expected contribution
from SM processes.
No significant excess of events has been found with respect to the SM 
prediction. 
Figure \ref{fig:hzz4l}\subref{fig:hzz4l-limit} shows the 95\% C.L. exclusion
limit on the production cross section for the SM Higgs boson as a function
of its mass. Details on the number of candidates selected in data and MC 
are illustrated in Table \ref{tab:hzztable}. 
A SM Higgs boson with mass $191<m_H<197$, $199<m_H<200$ and $214<m_H<224$ GeV 
is excluded. The observed
limits are consistent within less than 2$\sigma$ with the expected ones.
\begin{table}[ht]
\begin{center}
\begin{fntable}[0.6\columnwidth]
\begin{tabular}{c|c|c|c}
\hline
sample     & $4\mu$           & $2e 2\mu$        & $4e$    \\
           & \small{(L=2.28 fb$^{-1}$)} & \small{(L=1.96 fb$^{-1}$)} & \small{(L=1.98 fb$^{-1}$)}  \\
\hline

 observed  &  12         &  9            &  6              \\
 expected  & 8.8$\pm$1.2 & 11.1$\pm$1.4  &  4.32$\pm$0.52  \\
  
\hline
\end{tabular}
\end{fntable}
\caption{The observed and expected number of events after full selection
by the analysis for the $H\rightarrow ZZ^{(*)}\rightarrow 4l$ search
\cite{H4lATLAS}.
}
\label{tab:hzztable} 
\end{center} 
\end{table}
More details on this study are available in the paper recently submitted
\cite{H4lATLAS}.

\section{Overall SM Higgs Combination}

\begin{figure}[ht]
\begin{center}
\subfigure[]{
\includegraphics[width=0.490\columnwidth]{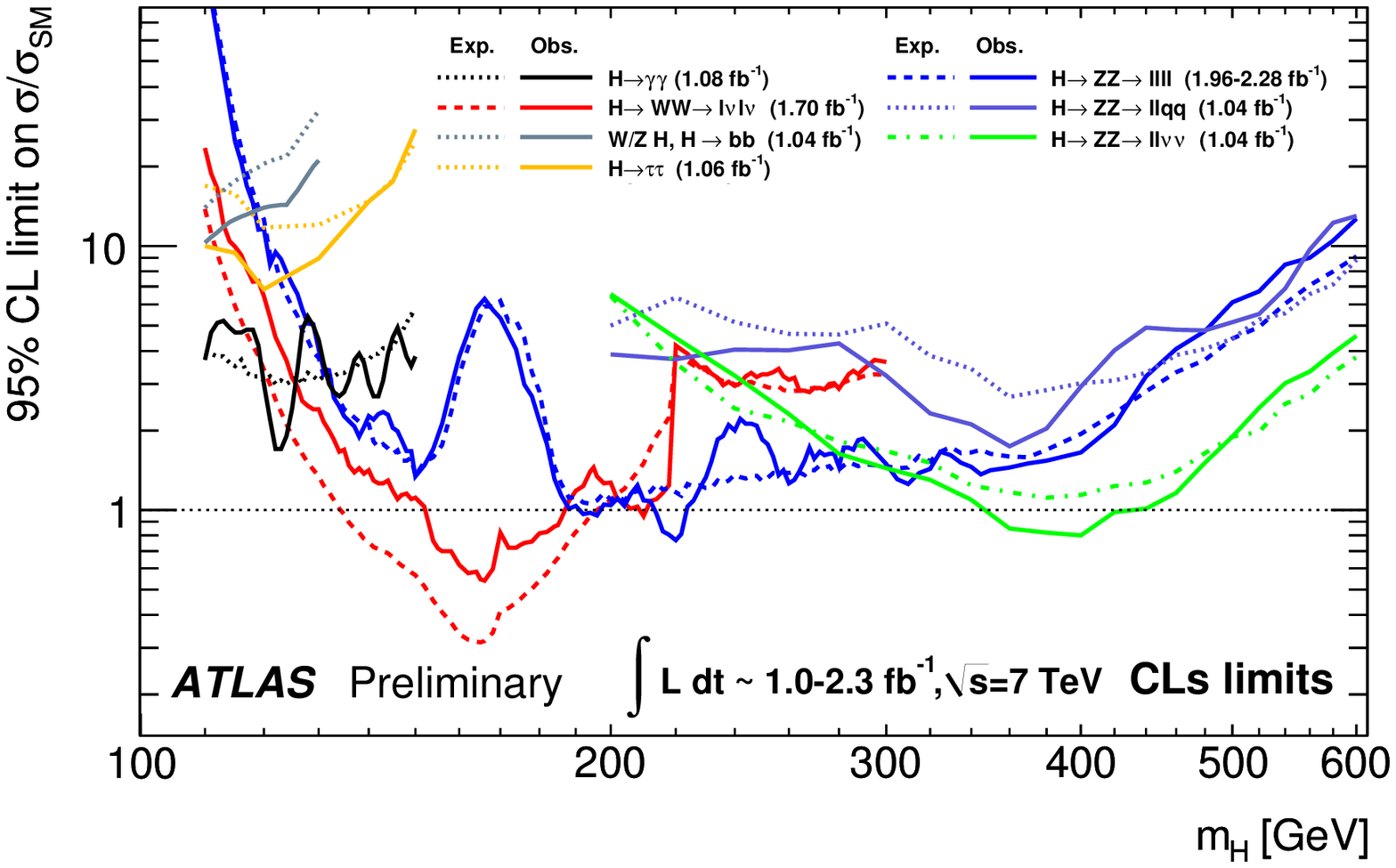}
\label{fig:hcombo-inputs}
}
\hspace{0.5cm}
\subfigure[]{
\includegraphics[width=0.425\columnwidth]{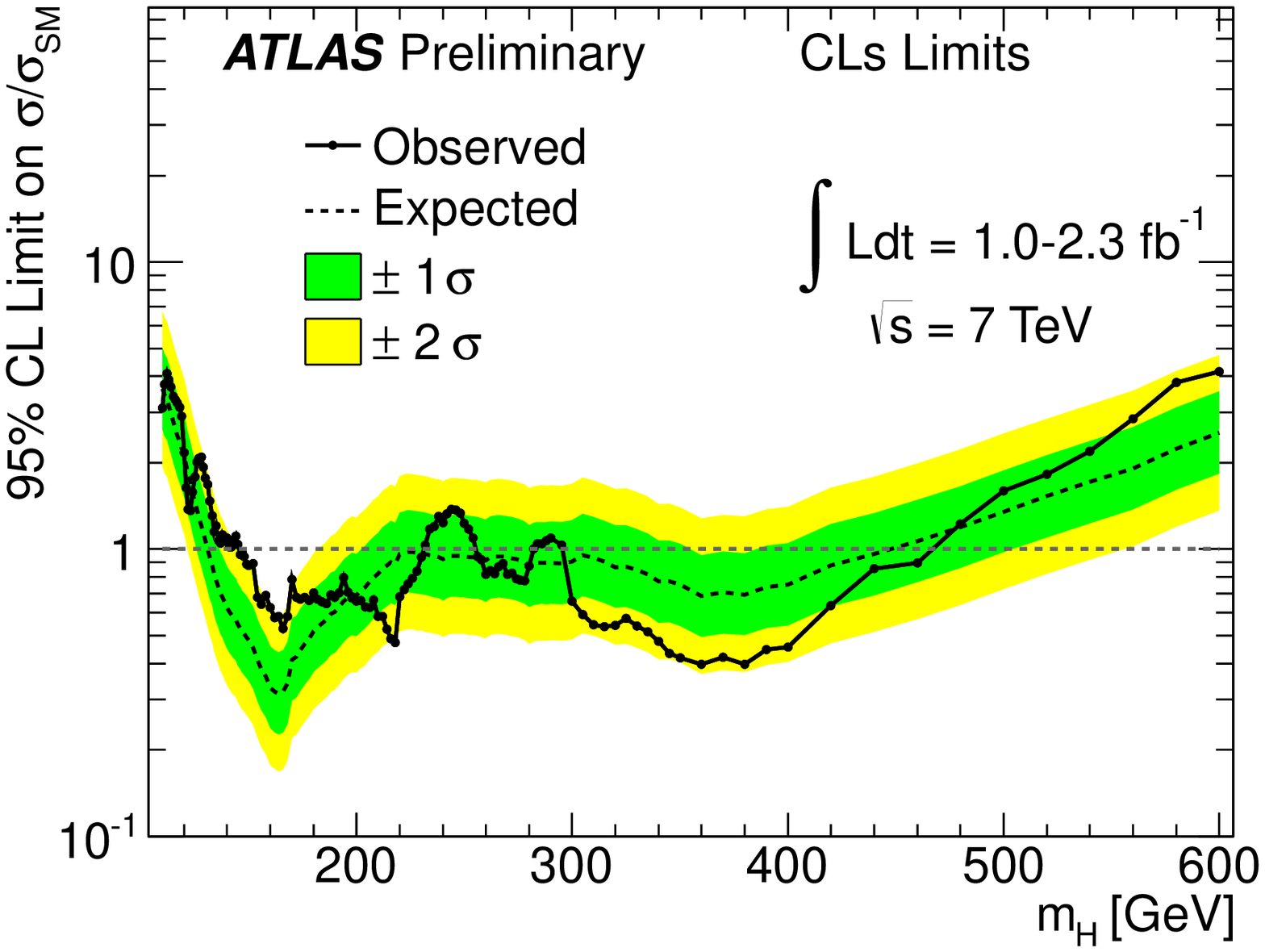}
\label{fig:hcombo-output}
}
\caption{
\subref{fig:hcombo-inputs} The expected (dashed) and observed (solid) 
cross section 95\% exclusions limits for the individual search channels, 
normalised to the Standard Model Higgs boson cross section, as functions 
of the Higgs boson mass. 
\subref{fig:hcombo-output} The combined upper limit on the Standard Model Higgs 
boson production cross section divided by the Standard Model expectation as a 
function of $m_H$ is indicated by the solid line
\cite{Hcomboresults}. 
}
\label{fig:hcombo}
\end{center}
\end{figure} 

All channels presented in this paper are combined with a statistical
procedure publicly available \cite{Hcombo}, which is again based on 
the modified frequentist method $CLs$ \cite{CLs}.
The profile likelihood ratio is used as a test statistics, whose 
distribution function is obtained
either with toy Monte Carlo pseudo-experiments, or using asymptotic
analytic expressions \cite{PL}.
The uncertainties of correlated experimental uncertainties (JES,
luminosity,...), across different Higgs channels, are taken into account. 
A careful treatment of theory uncertainties is also made, 
paying again attention to those that are correlated,
as for example the uncertainty on the Higgs boson production cross section
\cite{YR1}.
Figure \ref{fig:hcombo}\subref{fig:hcombo-inputs} summarizes the individual
standalone exclusion limits associated to the Higgs boson channels
discussed in this paper. The result of the combination of these results
is shown in figure Figure \ref{fig:hcombo}\subref{fig:hcombo-output}.
The production of a Standard Model Higgs boson is excluded at 95\% C.L.
in the mass regions $144<m_H<232$ GeV, $256<m_H<282$ GeV, and
$296<m_H<466$ GeV.
Moreover, the mass regions between 160 and 220 GeV, and between 300 
420 GeV, are excluded at least with 99

The exclusion limit in the mass region $110<m_H<135$ GeV is given mainly
from the combination of the $H\rightarrow \gamma\gamma$ and 
$H\rightarrow WW^{(*)}\rightarrow l\nu l\nu$ decay channels. For mass values
of $m_H$ up to 180 GeV, the limit is dominated by the 
$H\rightarrow WW^{(*)}\rightarrow l\nu l\nu$ decay, while the channels
$H\rightarrow ZZ\rightarrow 4l$ and $H\rightarrow ZZ\rightarrow ll\nu\nu$
dominate the exclusion limit for $m_H>180$ GeV.

The expected exclusion covers Standard Model Higgs boson mass range from 
131 GeV to 450 GeV.
The observed limit is in very good agreement with the background only 
expected limit over most of the studied mass range, except in the
low mass region $130< m_H < 170$ GeV where the observed limit is
consistent with 2$\sigma$ with the one expected in the hypothesis of no
signal. This is mainly due to the results coming from the search for  
$H\rightarrow WW^{(*)}\rightarrow l\nu l\nu$ production.
More details on this combination are available in this
ATLAS public note \cite{Hcomboresults}.

The full data set collected during the 2011 run, more than 5 fb$^{-1}$ for
ATLAS and CMS, will definitively help to better understand the data,
improve the physics analyses and to increase further the sensitivity
to the Higgs boson search to a wider mass interval.

%
%

\bibliographystyle{pramana}
\bibliography{nisati-arxiv}

\begin{thebibliography}{10}
\providecommand{\url}[1]{{\tt #1}}
\providecommand{\urlprefix}{URL }
\providecommand{\bibinfo}[2]{#2}
\providecommand{\eprint}[2][]{\url{#2}}

\bibitem{djouadi}
\bibinfo{author}{Abdelhak Djouadi}, {\em \bibinfo{journal}{Proceedings of the
  Lepton Photon 2011 Conference, {\normalfont to appear in} {\it Pramana -
  journal of physics}}\/} .

\bibitem{LEP}
\bibinfo{author}{LEP Working~Group for Higgs~boson searches}, {\em
  \bibinfo{journal}{Phys. Lett. B}\/} {\bf \bibinfo{volume}{565}},
  \bibinfo{pages}{61} (\bibinfo{year}{2003}).

\bibitem{verzocchi}
\bibinfo{author}{Marco Verzocchi}, {\em \bibinfo{journal}{Proceedings of the
  Lepton Photon 2011 Conference, {\normalfont to appear in} {\it Pramana -
  journal of physics}}\/} .

\bibitem{ATLAS}
\bibinfo{author}{ATLAS Collaboration}, {\em \bibinfo{journal}{JINST}\/} {\bf
  \bibinfo{volume}{3}}, \bibinfo{pages}{823} (\bibinfo{year}{2008}).

\bibitem{sharma}
\bibinfo{author}{Vivek Sharma}, {\em \bibinfo{journal}{Proceedings of the
  Lepton Photon 2011 Conference, {\normalfont to appear in} {\it Pramana -
  journal of physics}}\/} .

\bibitem{HggATLAS}
\bibinfo{author}{ATLAS Collaboration}, {\em
  \bibinfo{journal}{arXiv:1108.5895}\/}  (\bibinfo{year}{2011}).
  \bibinfo{note}{Accepted by Physics Letters B}.

\bibitem{HbbATLAS}
\bibinfo{author}{ATLAS Collaboration}, {\em
  \bibinfo{journal}{ATLAS-CONF-2011-103,
  http://cdsweb.cern.ch/record/1369826}\/} .

\bibitem{Htt1ATLAS}
\bibinfo{author}{ATLAS Collaboration}, {\em
  \bibinfo{journal}{ATLAS-CONF-2011-133,
  http://cdsweb.cern.ch/record/1383836}\/} .

\bibitem{Htt2ATLAS}
\bibinfo{author}{ATLAS Collaboration}, {\em
  \bibinfo{journal}{ATLAS-CONF-2011-132,
  http://cdsweb.cern.ch/record/1383835}\/} .

\bibitem{HWWlnln}
\bibinfo{author}{ATLAS Collaboration}, {\em
  \bibinfo{journal}{ATLAS-CONF-2011-134,
  http://cdsweb.cern.ch/record/1383837}\/} .

\bibitem{H4lATLAS}
\bibinfo{author}{ATLAS Collaboration}, {\em
  \bibinfo{journal}{arXiv:1109.5945s}\/}  (\bibinfo{year}{2011}).
  \bibinfo{note}{Accepted by Physics Letters B}.

\bibitem{HllnnATLAS}
\bibinfo{author}{ATLAS Collaboration}, {\em
  \bibinfo{journal}{arXiv:1109.3357}\/}  (\bibinfo{year}{2011}).
  \bibinfo{note}{Submitted to Physics Letters B}.

\bibitem{HllqqATLAS}
\bibinfo{author}{ATLAS Collaboration}, {\em
  \bibinfo{journal}{arXiv:1108.5064}\/}  (\bibinfo{year}{2011}).
  \bibinfo{note}{Submitted to Physics Letters B}.

\bibitem{HWWlnqq}
\bibinfo{author}{ATLAS Collaboration}, {\em
  \bibinfo{journal}{arXiv:1109.3615}\/}  (\bibinfo{year}{2011}).
  \bibinfo{note}{Accepted by Physics Review Letters}.

\bibitem{YR1}
\bibinfo{author}{S.~Dittmaier~{\it et al.,} LHC Higgs Cross Section
  Working~Group}, {\em \bibinfo{journal}{arXiv:1101.0593}\/}
  (\bibinfo{year}{2011}).

\bibitem{CLs}
\bibinfo{author}{A.~L. Read}, {\em \bibinfo{journal}{NJ. Phys. G}\/} {\bf
  \bibinfo{volume}{28}}, \bibinfo{pages}{2963} (\bibinfo{year}{2011}).

\bibitem{PL}
\bibinfo{author}{E.~Gross G.~Cowan, K.~Cranmmer} and
  \bibinfo{author}{O.~Vitells}, {\em \bibinfo{journal}{Eur. Phys. J.C}\/} {\bf
  \bibinfo{volume}{71}}, \bibinfo{pages}{1554} (\bibinfo{year}{2011}).

\bibitem{btagging}
\bibinfo{author}{ATLAS Collaboration}, {\em
  \bibinfo{journal}{ATLAS-CONF-2011-102,
  http://cdsweb.cern.ch/record/1369219}\/} .

\bibitem{CollApp}
\bibinfo{author}{M.~Soldate~J.J. Van der~Bij R.~K.~Ellis, I.~Hinchliffe}, {\em
  \bibinfo{journal}{Nucl. Phys. B}\/} {\bf \bibinfo{volume}{297}},
  \bibinfo{pages}{221} (\bibinfo{year}{1988}).

\bibitem{mmc}
\bibinfo{author}{A.~Pranko A.~Safanov A.~Elagin, P.~Murat}, {\em
  \bibinfo{journal}{Nucl. Instrum. Meth. A}\/} {\bf \bibinfo{volume}{654}},
  \bibinfo{pages}{481} (\bibinfo{year}{2011}).

\bibitem{Hcomboresults}
\bibinfo{author}{ATLAS Collaboration}, {\em
  \bibinfo{journal}{ATLAS-CONF-2011-135,
  http://cdsweb.cern.ch/record/1383838}\/} .

\bibitem{Hcombo}
\bibinfo{author}{ATLAS} and \bibinfo{author}{CMS Coll.}, {\em
  \bibinfo{journal}{ATL-PHYS-PUB-2011-011,
  http://cdsweb.cern.ch/record/1375842}\/} .

\end{thebibliography}

%
%

\end{document}